\documentclass[varenna]{cimento}

\usepackage{graphicx}  

\title{Measuring the mass profile of galaxy clusters beyond their virial radius}

\author{Antonaldo Diaferio}

\institute{Dipartimento di Fisica Generale ``Amedeo Avogadro'', Universit\`a degli Studi
di Torino\\ Via P. Giuria 1, I-10125 Torino, Italy}

\institute{Istituto Nazionale di Fisica Nucleare, Sezione di Torino,\\ 
Via P. Giuria 1, I-10125 Torino, Italy}

\begin{document}

\maketitle

\begin{abstract}
Traditional estimators of the mass of galaxy clusters assume 
that the cluster components (galaxies, intracluster medium, and dark matter) are in dynamical
equilibrium. Two additional
estimators, that do not require this assumption, were proposed in the 1990s:
 gravitational
lensing and the caustic technique. With these methods, we can measure the cluster mass
within radii much larger than the virial radius. In the caustic technique, 
the mass measurement is only based on the celestial
coordinates and redshifts of the galaxies in the cluster field of view; therefore, 
unlike lensing, it can be, in principle, applied to clusters at any redshift. 
Here, we review the origin, the basics and the performance of the caustic method.
\end{abstract}

\section{Introduction}

In the currently accepted cosmological model, galaxy formation 
is intimately connected to the formation of the large-scale cosmic structure. 
To test this model, we need to measure the relative distribution of light and matter
in the Universe. The mass distribution on small scales, from galaxies to galaxy
clusters, has been usually inferred by assuming that systems are in dynamical
equilibrium. On very large scales, the mass overdensities are
small enough that the linear theory of density perturbations can be
used to measure the mass distribution from the relation between the mass density 
field and the peculiar velocities of galaxies \cite{zaroubi02}.

On intermediate, mildly non-linear scales, $\sim 1-10h^{-1}$~Mpc,\footnote{We use 
$H_0=100 h$ km s$^{-1}$ Mpc$^{-1}$ throughout.} neither
the dynamical equilibrium hypothesis nor linear theory are valid.
No robust way of measuring the mass distribution in 
this regime was available until the 1990s, when both
gravitational
lensing and the caustic technique were developed.  
Here, we provide an overview of how the caustic technique
came about and what it has contributed so far.

\section{The context of mass estimators}

\subsection{The assumption of dynamical equilibrium}

Galaxy cluster mass estimators measure either the  total mass within a given
radius $R$ or the mass radial profile. Traditionally, both
kinds of estimators are based on the
assumption that the cluster is spherical and in dynamical equilibrium.

The virial theorem is usually applied when the number of
galaxy redshifts is not large: the galaxy velocity dispersion $\sigma$
and the cluster size $R$ are sufficient to yield an estimate of
the cluster total mass $M=\sigma^2 R/G$ \cite{zwicky37},
where $G$ is the gravitational constant. More accurate measurements require
a surface term correction \cite{the86, merritt87}, which can decrease
the estimated mass by a substantial factor ($\sim 20\%$, on average \cite{girardi98}),
and knowledge of the galaxy orbital distribution;
however, although this distribution can only be reasonably guessed in
most cases, its uncertainty has only a modest impact ($\sim 5\%$) on the final mass estimate.
These uncertainties become an order of magnitude larger if the galaxies are
not fair tracers of the mass distribution.

When the number of galaxy spectra is large enough that we can estimate the
velocity dispersion profile, we can apply the Jeans equations for a steady-state
spherical system.  The cumulative mass is
\begin{equation}
M(<r) = - {\langle v_r^2 \rangle r \over G} \left[{{\rm d}\ln\rho_{\rm m}\over {\rm d}\ln r} + 
{{\rm d}\ln\langle v_r^2 \rangle\over {\rm d}\ln r} 
+ 2\beta(r)\right] \; .
\label{eq:jeans}
\end{equation}
However, as in the virial theorem, 
the application of equation (\ref{eq:jeans}) requires the assumption of a relation between 
the galaxy number density profile and the mass density profile $\rho_{\rm m}$. Moreover,
we do not usually know the velocity anisotropy parameter 
\begin{equation}
\beta(r)=1-{\langle v^2_\theta\rangle + \langle v^2_\phi\rangle\over 2\langle v^2_r\rangle} \;  ,
\label{eq:beta}
\end{equation}
where $v_\theta$, $v_\phi$, and $v_r$ are the longitudinal, azimuthal and
radial components of the velocity $v$ of the galaxies, respectively, and the brackets
indicate an average over the velocities of the galaxies in the volume ${\rm d}^3{\bf r}$
centered on position ${\bf r}$ from the cluster center. Therefore, 
we can not measure $M(<r)$ without guessing $\beta(r)$, or vice versa. 
A common strategy is to measure the velocity distributions of different galaxy populations
which are assumed to be in equilibrium and thus to trace the same
gravitational potential. This method can help to break this mass-anisotropy degeneracy,
although not completely (see \cite{biviano06, biviano08} for
very lucid reviews of these methods).

We can estimate the mass profile when observations in the X-ray band provide
the intracluster medium (ICM) density $\rho_{\rm gas}$ and temperature $T$.
The assumption of hydrostatic equilibrium of the ICM 
yields a relation similar to equation (\ref{eq:jeans}) 
\begin{equation}
M(<r) = - {k T r\over G\mu m_{\rm p}} \left[{{\rm d}\ln\rho_{\rm gas}\over {\rm d}\ln r} + {{\rm d}\ln T\over {\rm d}\ln r}
\right] \; 
\label{eq:ICM}
\end{equation}
where $k$ is the Boltzmann constant, $\mu$ the mean molecular weight, and $m_{\rm p}$
the proton mass. Note that the term analogous to $\beta$, 
which appears in equation (\ref{eq:beta}), 
is now zero, because, unlike the galaxy orbits, the ICM pressure is isotropic.
When  a sufficient angular resolution and energy sensitivity are not available to measure
the X-ray spectrum at different clustrocentric radius and thus estimate
the temperature profile, an isothermal ICM is usually assumed. However, 
the departure from this assumption appears to be substantial in most clusters
where the density and temperature structures can be measured (e.g., \cite{mark98, deg02}).

For estimating the cluster mass when detailed observations of the cluster are unavailable, 
we can use a scaling relation between the mass and an observable average quantity.
The most commonly used scaling relations are those involving 
ICM thermal properties, as the X-ray temperature (e.g. \cite{pierpaoli03}; see also
\cite{borgani06, borgani08} for reviews). 
In this case, however, the complex thermal structure of the ICM
can significantly bias the cluster mass estimate \cite{rasia06}.
Rather than using an X-ray observable, one could use, in principle, the integrated
Sunyaev-Zel'dovich effect, which yields a correlation with mass 
which is tighter than the mass-X-ray temperature
correlation \cite{motl05}. However, this correlation is currently valid only 
for simulated clusters,
and still needs to be confirmed by upcoming cluster surveys.

\subsection{Dropping the dynamical equilibrium assumption}

The astrophysical
relevance of galaxies as gravitational lenses was first intuited by Zwicky \cite{zwicky37},
but it was only fifty years later that the first gravitational lens effect was measured
in a galaxy cluster \cite{lyn86}.
The lensing effect is a distortion of the optical images of sources beyond
the mass concentration; this distortion depends only
on the amount of mass along the line-of-sight and not on the dynamical state of this mass.
The obvious advantage is thus that the dynamical equilibrium assumption,
that is essential for all the methods listed above, becomes now unnecessary.
The lensing effect can be classified as strong or weak lensing,
depending on its intensity. Strong lensing creates multiple images
of a single source and can be used for measuring the cluster mass in its core,
where the gravitational potential is deep enough. In the outer regions, the
lensing effect is weaker and it only yields a tangential distortion of the induced ellipticities
of the shape of the background galaxies; weak lensing can thus 
measure the depth of the potential well from the center to the cluster outskirts.
The most serious disadvantage of gravitational lensing for measuring masses
is that the  signal intensity depends 
on the relative distances between observer, lens 
and source, and not all the clusters can clearly be in the appropriate position to provide
lens effects that are easily measurable. Moreover, weak lensing
does not generally have a large signal-to-noise
ratio and weak lensing analyses are not trivial (see e.g., \cite{schnei06}). 

In 1997, Diaferio and Geller \cite{diaf97} 
proposed the caustic technique, a novel method to estimate the cluster mass 
which is not based on the dynamical equilibrium assumption and
only requires galaxy celestial coordinates and redshifts. 
The method can thus measure the cluster mass on all
the scales from the central region to well beyond the virial radius $r_{200}$, 
the radius within which the average mass density is
200 times the critical density of the Universe. 
Prompted by the $N$-body simulations
of van Haarlem and van de Weygaert \cite{haarlem93b}, Diaferio and Geller \cite{diaf97} noticed
that in hierarchical models of structure
formation, the velocity field in the regions surrounding the cluster is not
perfectly radial, as expected in the spherical infall model \cite{regos89, hiotelis01},
but has a substantial random component. This fact can be exploited 
to extract the escape velocity of galaxies from their distribution 
in redshift space. Here, we will provide an overview of this method.

\subsection{Masses on different scales}

It is clear that weak lensing and the caustic technique can be applied
to scales larger than the virial radius because they do not depend on the 
assumption of dynamical equilibrium. However, the other estimators we mentioned above
do not always measure the total cluster mass within $r_{200}$,
as, for example, the virial analyses, based on optical observations, usually do.
X-ray estimates rarely go beyond $\sim 0.5 r_{200}$, because on
these larger scales the X-ray surface brightness
becomes smaller than the X-ray telescope sensitivity; gravitational lensing only 
measures the central mass within $\sim 0.1 r_{200}$, where the strong regime 
applies. Of course, scaling relations do not provide any information on the mass profile,
but they rather provide the total mass within a given radius which depends on the scaling
relation used: typically, X-ray, optical and Sunyaev-Zel'dovich scaling relations 
yield masses within increasing radii, but still smaller than $r_{200}$.

\section{History}

The spherically symmetric infall onto an initial density perturbation is
the simplest classical problem we encounter when we treat the formation
of cosmic structure by gravitational instability in an expanding background 
\cite{gunn72, bert85}. The solution to this problem provides two relevant results: 
the density profile of the resulting system and the mean
mass density of the Universe $\Omega_0$. 

The former issue has a long history that we do not review here
(see, e.g., \cite{zaroubi96, delpop04}). 
The basic idea is simple: we can imagine a spherical perturbation 
separated into individual spherical mass shells
that expand to the maximum turn-around radius, the radius where
the radial velocity $v_{\rm pec}(r)$ equals the Hubble velocity, before 
starting to collapse.
This simple picture enables us to predict the density profile of the
final object if we assume that mass is conserved, there is no shell crossing, 
and we know the initial density profile of the perturbation, namely
the initial two-point mass correlation function $\xi(r)$; $\xi(r)$ 
contains the same information as the power spectrum $P(k)$ of the mass density 
perturbations, if these are Gaussian variates.
For scale-free initial power spectra $P(k)\propto k^n$, the final density profile is
$\rho\propto r^{-\alpha}$ with
$\alpha$ depending on $\Omega_0$ and $n$. 

The spherically symmetric infall can also be used to estimate $\Omega_0$. 
When the average mass overdensity $\delta(r)$ within the radius $r$ 
of the perturbation is small enough, we can compute the radial velocity of each shell of
radius $r$ according to linear theory 
\begin{equation}
{v_{\rm pec}(r)\over H_0 r} =-{1\over 3}\Omega_0^{0.6}\delta(r)\; .
\label{eq:vpec-lin}
\end{equation}
In the simplest application of this relation to real systems, we assume that galaxies trace mass, 
so that the galaxy number overdensity is simply related
to $\delta$; a measure of ${v_{\rm pec}}$ thus promptly yields $\Omega_0$.
In the 1980s this strategy was applied to the Virgo cluster and the Local Supercluster; 
galaxies in these systems 
are close enough that we can measure galaxy distances $d$ independently of redshift $cz$,
and thus estimate the projection along the line of sight, $v_{\rm pec}^{\rm los}=cz-H_0d$, 
of the radial velocity $v_{\rm pec}$. 
These analyses indicated that $\Omega_0=0.35\pm 0.15$ \cite{davis83}, 
in agreement with the most recent estimates \cite{dunkley08}, but at odds with 
the inflationary value $\Omega_0=1$, which, at that time, was 
commonly believed to be the ``correct'' value.

A slight complication derives from the fact that the external regions of 
clusters are not properly described by linear 
theory. We can use instead the spherical infall model. In this case, 
$\delta$ and $\Omega_0$ are still separable quantities
and we can recast equation (\ref{eq:vpec-lin}) as
\begin{equation}
{v_{\rm pec}(r)\over H_0 r} = -{1\over 3}\Omega_0^{0.6}f(\delta)
\label{eq:vpec-nonlin}
\end{equation}
so that we can still measure $\Omega_0$ once $\delta$ is known.  Typical approximations are
$f(\delta)=\delta(1+\delta)^{-1/4}$ \cite{yahil85} and $f(\delta)=\delta(1+\delta/3)^{-1/2}$ 
\cite{vill86, cupani08}. A more serious complication is that departures
from spherical symmetry can be large in real systems 
and the radial velocities $v_{\rm pec}$ derived
from their line-of-sight components can be affected by relative uncertainties 
of the order of 50\% \cite{vill86}. 

The measure of absolute distances to galaxies remains a difficult problem today. 
Thus, estimating $\Omega_0$ from the infall regions of clusters might not be
trivial. However, this complication  
can be bypassed by the intuition of Reg\"os and Geller \cite{regos89}
who were inspired by the work of Shectman \cite{shect82},  
Kaiser \cite{kais87} and Ostriker {\it et al.} 
\cite{ostr88}. Kaiser showed that
when observed in redshift space (specifically the line-of-sight velocities of galaxies $cz$ 
versus their clustrocentric angular
distance $\theta$), the infall pattern around a rich cluster appears as a ``trumpet horn''
whose amplitude ${\cal A}(\theta)$ decreases with $\theta$.
The turn-around radius is identified by the condition ${\cal A}(\theta)=0$ \cite{ostr88}. 
For the Abell cluster A539, Ostriker {\it et al.} \cite{ostr88} 
found the turn-around radius $\theta_{\rm ta}\sim 2^\circ\sim 3 h^{-1}$~Mpc. 
Although the galaxy sampling in the infall region of this cluster was too sparse to measure $\Omega_0$
(they only set a lower limit $\Omega_0>0.03$ with equation \ref{eq:vpec-nonlin}), 
the proposed strategy was intriguing, because it was showed that
measuring
galaxy distances independently of redshift was unnecessary.

\begin{figure}
\centering
\includegraphics[width=0.8\textwidth]{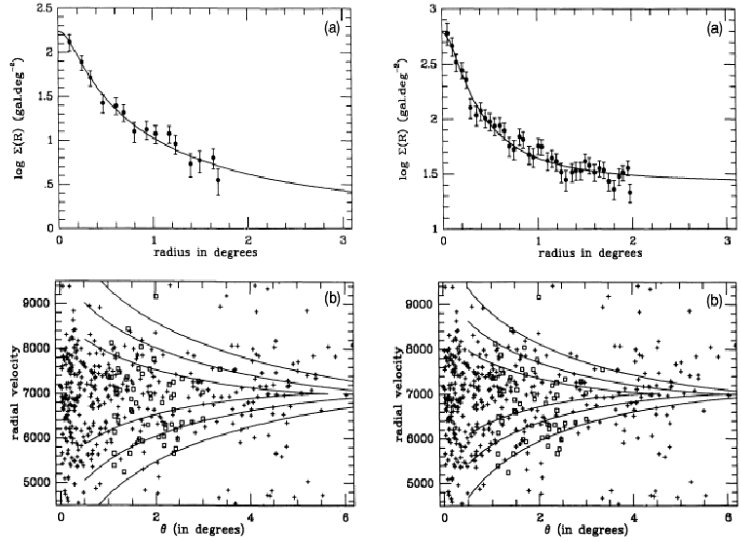}
\caption{Caustics (solid lines) according to the spherical infall model (equation \ref{eq:regos}) in the Coma cluster
(lower panels). The symbols show the galaxy positions in the redshift diagram. 
Larger amplitudes correspond to increasing cosmic densities $\Omega_0=0.2$, $0.5$, $1.0$. 
The mass overdensity $\delta$ 
is estimated from the galaxy number densities (upper panels) based on CfA data (left panels), or APM data 
(right panels). From \cite{haarlem93}.}
\label{fig:vanhaarl93}
\end{figure}

\begin{figure}
\centering
\includegraphics[angle=-90,width=0.8\textwidth]{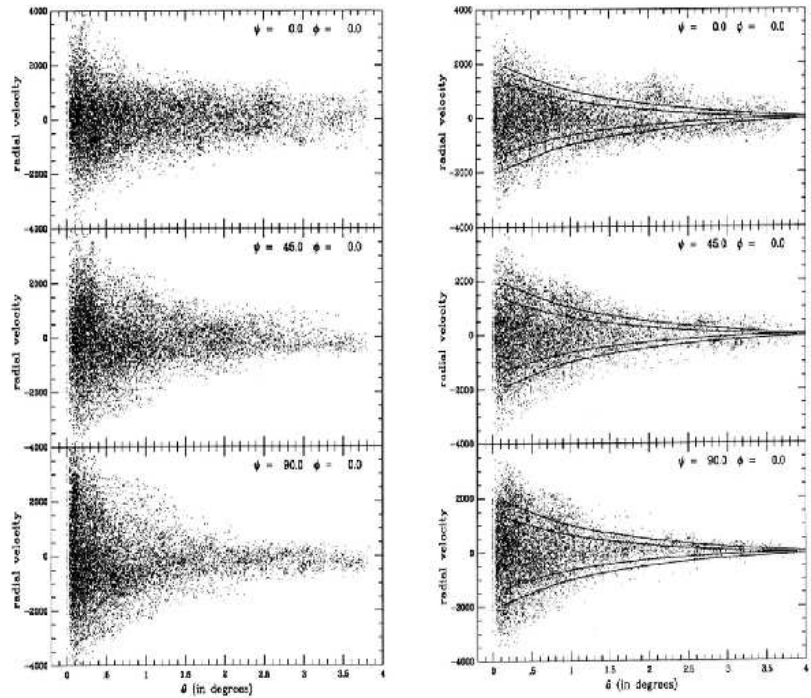}
\caption{Redshift diagrams of clusters
in an $N$-body simulation of the standard Cold Dark Matter (CDM) 
model with $\Omega_0=1$. The dots show the dark matter particle positions. The left column
shows the redshift diagrams of the same cluster, observed along three
different lines of sight, 
that has just accreted a group.
The right column shows the redshift diagram of another cluster, observed along three
different lines of sight, that has not 
had substantial mass accretion in the recent past. 
In the right column, the caustics according to the spherical infall model
are also shown as solid lines; the smaller (larger) amplitude 
corresponds to $\Omega_0=0.5$, ($\Omega_0=1$), whereas $\Omega_0=1$ in the simulation. From \cite{haarlem93b}.}
\label{fig:vanhaarl93b}
\end{figure}

Reg\"os and Geller \cite{regos89} quantified this intuition by
showing that the relation between the galaxy number density $\bar n(r)$ in real
space and the galaxy number density $n(cz,\theta)$ in redshift space
is 
\begin{equation}
n(cz,\theta)=\bar n(r)\left(r\over cz\right)^2 {1\over J}
\end{equation}
where $J$ is the Jacobian of the transformation from real to redshift
space coordinates. When $J=0$, $n(cz,\theta)$ is infinite. This condition locates the borders
of Kaiser's horn which are named {\it caustics}. 
We can now use equation  (\ref{eq:vpec-nonlin}) to relate ${\cal A}(\theta)$ to $\Omega_0$
(equation 34 of \cite{regos89}):
\begin{equation}
{\cal A}(\theta)\sim \Omega_0^{0.6} rf(\delta) \left[
-{{\rm d}\ln f(\delta)\over {\rm d}\ln r}\right]^{-1/2}
\label{eq:regos}
\end{equation} 
where $r$ and $\theta$ are related by the transformation from real to redshift
space coordinates. 

\begin{figure}
\centering
\includegraphics[angle=-90,width=0.7\textwidth]{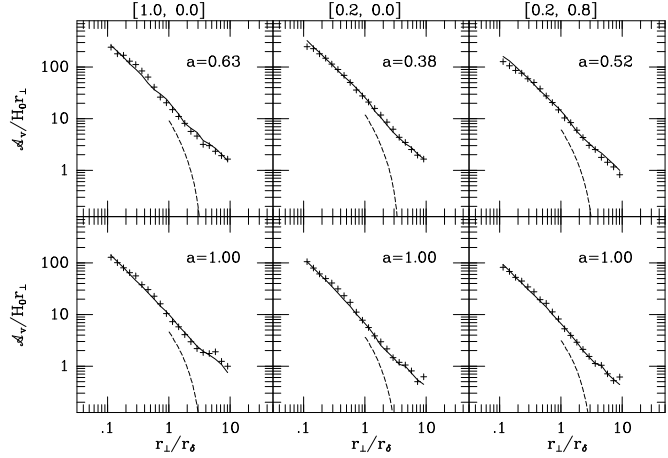}     
\caption{Caustic amplitude vs. projected clustrocentric distance for a simulated cluster
in three different CDM cosmologies (columns) with $[\Omega_0,\Omega_\Lambda]$ as shown
above the upper panels. The cluster is shown right after a major merger (upper row)
and at equilibrium (lower row). The cosmic time is shown by the scale factor $a$.
The crosses show the actual caustic amplitude. The solid lines show 
the line-of-sight component of the square of the escape velocity: 
$\langle v^2_{\rm esc, los}(r)\rangle^{1/2} = \{- 2\phi(r)[1-\beta(r)]/[3-2\beta(r)]\}^{1/2}$. The dashed lines show 
the prediction of the spherical infall model. Clustrocentric distances 
are in units of the virial radius $r_\delta$. From \cite{diaf97}.}
\label{fig:vesc}
\end{figure}

Unfortunately, the caustics appeared to be very fuzzy when 
a sufficiently dense sampling of the infall region of 
a rich cluster like Coma  was obtained \cite{haarlem93};
consequently, the measure
of $\Omega_0$ was rather uncertain (Figure \ref{fig:vanhaarl93}).
This disappointing result was attributed to the fact that the assumption of spherical symmetry 
is very poorly satisfied and that the substructure surrounding the cluster 
distorts the radial velocity field \cite{haarlem93b} (Figure \ref{fig:vanhaarl93b}).
Being so sensitive to the cluster shape, locating the caustics 
in the redshift diagram did not appear to be a promising strategy to measure $\Omega_0$.  

A breakthrough came when Diaferio and Geller \cite{diaf97}
took a step further than van Haarlem and van de Weygaert. 
In hierarchical clustering scenarios, 
clusters accrete mass episodically and anisotropically \cite{colberg99} 
rather than through the gentle infall of spherical shells.
Moreover, clusters accrete galaxy groups with their own 
internal motion. Therefore, the velocity field of the infall region 
can have substantial non-radial and random components.
These velocity components both make the
caustic location fuzzy, and, more importantly, increase the caustic amplitude
when compared to the spherical infall model.

This intuition opened the way to interpret the square of the caustic amplitude ${\cal A}^2(\theta)$ as
the average, over the volume ${\rm d}^3{\bf r}$, 
of the square of the line-of-sight component of the escape velocity 
$\langle v^2_{\rm esc, los}(r)\rangle =\left[-2\phi(r) g^{-1}(\beta)\right]^{1/2}$, where
$\phi(r)$ is the gravitational potential profile 
and $g$ (equation \ref{eq:gbeta}) is a function 
of the velocity parameter anisotropy $\beta(r)$. 
The crucial point here is that the equation ${\cal A}^2(r) = \langle v^2_{\rm esc, los}(r)\rangle $
holds {\it independently of the dynamical state of the cluster}.

This interpretation works amazingly well. Figure \ref{fig:vesc} shows
the results of $N$-body simulations of the formation and evolution of a 
galaxy cluster in Cold Dark Matter (CDM) models with 
different cosmological parameters.
The caustic amplitude (crosses) and $\langle v^2_{\rm esc, los}(r_\perp)\rangle $ (solid lines),
as a function of the projected distance $r_\perp$, agree
at all scales out to ten virial radii $r_\delta$\footnote{See \cite{diaf97} for the
proper definition of the virial radius $r_\delta$ in these plots.} and independently of the dynamical
state of the cluster: immediately after a
major merging (upper panels) or at equilibrium (lower panels).
The spherical infall model (dashed lines), which should only hold for $r_\perp>r_\delta$, 
always severely underestimates the actual caustic amplitude.
These simulations and those in \cite{diaf99} also show another
relevant result: the major effect of the cluster shape is
not to make the caustics fuzzy but rather to yield different 
caustic amplitudes depending on the line of sight.

The identification ${\cal A}^2(r) = \langle v^2_{\rm esc, los}(r)\rangle $
 can be immediately used to measure the cluster mass.
If we assume spherical symmetry, the cumulative total mass $M(<r)$ is
\begin{equation}
GM(<r) = r^2{{\rm d}\phi\over {\rm d}r} =  
-{r\over 2} \langle v^2_{\rm esc, los}\rangle g(\beta)\left({{\rm d}\ln \langle v^2_{\rm esc, los}\rangle \over {\rm d}\ln r} 
+ {{\rm d}\ln g \over {\rm d}\ln r}\right) \; .
\label{eq:dphi}
\end{equation}
However, in realistic situations, the two logarithmic derivatives are comparable,
and we thus need to know $\beta(r)$: this is generally not the case.
Moreover, the most serious obstacle in using equation (\ref{eq:dphi}) is the
fact that sparse sampling and background and foreground
galaxies yield the estimate of $\langle v^2_{\rm esc, los}(r)\rangle$ too 
noisy to extract accurate information from its differentiation. 

To bypass this problem, Diaferio and Geller \cite{diaf97} suggested a different recipe
to estimate the cumulative mass
\begin{equation}
GM(<r)={\cal F}_\beta\int_0^r \langle v^2_{\rm esc, los}(r)\rangle {\rm d}r =  {\cal F}_\beta\int_0^r {\cal A}^2(r) {\rm d}r 
\end{equation}
where ${\cal F}_\beta\approx 0.5$ is a constant.
This recipe has been applied to a large number of clusters ever since and 
it is now becoming a popular tool to measure the mass in the cluster infall regions. 
Below, we justify this recipe and show how it works in practice.

\section{The caustic method}

In hierarchical clustering models of structure formation, clusters form
by the aggregation of smaller systems accreting onto the cluster from
the surrounding region. The accretion does not happen purely radially
and galaxies within the falling clumps have
velocities with substantial non-radial components. Specifically, 
these velocities depend both on the tidal fields
of the surrounding region and on the gravitational potential
of the clusters and the groups where the galaxies reside.
In the previous section, we have seen that, when viewed in the redshift diagram, galaxies populate
a region with a characteristic trumpet shape whose amplitude, which decreases 
with increasing $r$,  is related to the escape
velocity from the cluster region.

The escape velocity $v_{\rm esc}^2(r)=-2\phi(r)$, where $\phi(r)$ is the
gravitational potential originated by the cluster, is a non-increasing function of $r$,
because gravity is always attractive and ${\rm d}\phi/{\rm d}r>0$. Thus,
we can identify the square of the amplitude
${\cal A}$ at the projected radius $r_\perp$ as the average of the square of the 
line-of-sight component $\langle v^2_{\rm los}\rangle$ of the escape velocity
at the three-dimensional radius $r=r_\perp$. To relate $\langle v^2_{\rm los}\rangle$ 
to $\phi(r)$, we need the velocity anisotropy parameter $\beta(r)$ (equation \ref{eq:beta}).  
If the cluster rotation is negligible, we have
$\langle v^2_\theta\rangle=\langle v^2_\phi\rangle=\langle v^2_{\rm los}\rangle$, 
and $\langle v^2_r\rangle=\langle v^2\rangle-2\langle v^2_{\rm los}\rangle$. By substituting
this relation into equation (\ref{eq:beta}), 
we obtain $\langle v^2\rangle=\langle v^2_{\rm los} \rangle g(\beta)$ where
\begin{equation}
g(\beta) = {3-2\beta(r)\over 1-\beta(r)}\; .
\label{eq:gbeta}
\end{equation}
By applying this relation
to the escape velocity at radius $r$, $\langle v_{\rm esc}^2(r)\rangle=-2\phi(r)$,
and by assuming that ${\cal A}^2(r)=\langle v^2_{\rm esc, los}\rangle$,
we obtain the fundamental relation between the gravitational potential $\phi(r)$ and
the observable caustic amplitude ${\cal A}(r)$
\begin{equation}
-2\phi(r)={\cal A}^2(r)g(\beta) \; .
\label{eq:rig-pot}
\end{equation}

To infer the cluster mass to very large radii, 
one first notices that the mass of a shell of infinitesimal thickness ${\rm d}r$ can be
cast in the form $G{\rm d}m=-2\phi(r){\cal F}(r){\rm d}r = {\cal A}^2(r)g(\beta) {\cal F}(r) {\rm d}r$ where
\begin{equation}
{\cal F}(r)=-2\pi G{\rho(r)r^2\over \phi(r)}\; .
\end{equation}
Therefore the mass profile is
\begin{equation}
GM(<r)=\int_0^r {\cal A}^2(r) {\cal F}_\beta(r) {\rm d}r
\label{eq:rig-massprof}
\end{equation}
where ${\cal F}_\beta(r) = {\cal F}(r) g(\beta)$.

Equation (\ref{eq:rig-massprof}) however only relates the
mass profile to the density profile of a spherical system and one profile
can not be inferred without knowing the other. We can solve this
impasse by noticing that, in hierarchical clustering
scenarios, ${\cal F}(r)$ is not a strong function of $r$ \cite{diaf97}. This is easily seen
in the case of the Navarro, Frenk and White (NFW) \cite{nfw} mass density profile, 
which is an excellent description of the dark matter
distribution in these models:
\begin{equation}
{\cal F}_{\rm NFW}(r) = {r^2\over 2(r+r_s)^2}{1\over \ln(1+r/r_s)} \; 
\end{equation}
where $r_s$ is a scale-length parameter.
If clusters form through hierarchical clustering, ${\cal F}_\beta(r)$ is also a slowly 
changing function of $r$ \cite{diaf97, diaf99}.
We can then assume, somewhat strongly, that ${\cal F}_\beta(r)=
{\cal F}_\beta={\rm const}$
altogether and adopt the recipe
\begin{equation}
GM(<r)={\cal F}_\beta\int_0^r {\cal A}^2(r) {\rm d}r \; .
\label{eq:recipe-massprof}
\end{equation}
When ${\cal F}_\beta=1/2$, this recipe proves to yield mass profiles accurate to
50\% or better both in $N$-body simulations and in real clusters, when compared
with masses obtained with standard methods, namely Jeans equation, X-ray and gravitational lensing,
applied on scales where the validities of these methods overlap \cite{diaf05}.

It is appropriate to emphasize that equations (\ref{eq:rig-pot}) and (\ref{eq:rig-massprof}) are
rigorously correct, whereas
equation (\ref{eq:recipe-massprof}) is a heuristic recipe for the estimation of the mass profile.

\begin{figure}
\centering
\includegraphics[angle=90,width=0.5\textwidth]{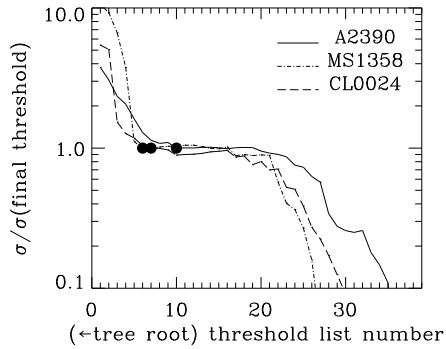}
\caption{Velocity dispersion of the galaxies on the main branch of
the binary tree of three real clusters
while walking towards the leaves (see \cite{diaf05}). 
There is an obvious plateau when entering the tree sector with
the cluster members. The filled dots indicate the chosen $\sigma$ used
to cut the binary tree and thus select the cluster members.}
\label{fig:thresholds}
\end{figure}

\subsection{Implementation}

The implementation of the caustic method requires: (1) the determination of the cluster
center; (2) the estimate of the galaxy distribution in the redshift diagram;
(3) the location of the caustics.

For estimating the cluster center, the galaxies in the cluster field
of view\footnote{We clarify that to apply the caustic technique we
already need to know that there is a cluster in the field of view.
The caustic technique, as it is currently conceived, is not a method to identify
clusters in redshift surveys, as the Voronoi tessellation \cite{ram01} or the matched filter
 \cite{post96}.} 
are arranged in a binary tree according to the pairwise
projected energy
\begin{equation}
E_{ij}=-G{m_i m_j\over R_p}+{1\over 2}{m_i m_j\over m_i+m_j}\Pi^2 
\label{eq:pairwise-energy}
\end{equation}
where $R_p$ and $\Pi$ are the projected spatial separation and the 
proper line-of-sight velocity difference of each galaxy pair respectively;
$m_i$ and $m_j$ are the galaxy masses which are usually set constant, 
but can also be chosen according to the galaxy luminosities.

By walking along the main branch of the tree from the root to the leaves,
we progressively remove the background and foreground galaxies. We  
identify the cluster members by computing the velocity dispersion $\sigma$ of the galaxies still on the main
branch at each step: $\sigma$ remains roughly constant when we move
through the binary tree sector which only contains the cluster member (Figure \ref{fig:thresholds}), 
because the cluster is approximately isothermal. 

The cluster members provide the cluster center and therefore the redshift
diagram $(r,v)$. The galaxy distribution $f_q(r,v)$ on this plane is estimated
with an adaptive kernel method. At each projected radius $r$, the function 
$\varphi(r)=\int f_q(r,v) dv$ provides the mean escape velocity
$\langle v_{\rm esc}^2\rangle_{\kappa,R}=\int_0^R{\cal A}_\kappa^2(r)\varphi(r)dr/
\int_0^R\varphi(r)dr$ where ${\cal A}_\kappa$ is the amplitude of the 
caustics located by the equation $f_q(r,v)=\kappa$. The appropriate $\kappa$
is the root of the equation $\langle v_{\rm esc}^2\rangle_{\kappa,R}=4\sigma^2$,
where $\sigma^2$ is the velocity dispersion of the members identified
on the binary tree. 
Further technical
details of this implementation are described in \cite{diaf99, serra}.

\section{Reliability of the method}

\begin{figure}
\centering
\includegraphics[angle=-90,width=0.7\textwidth]{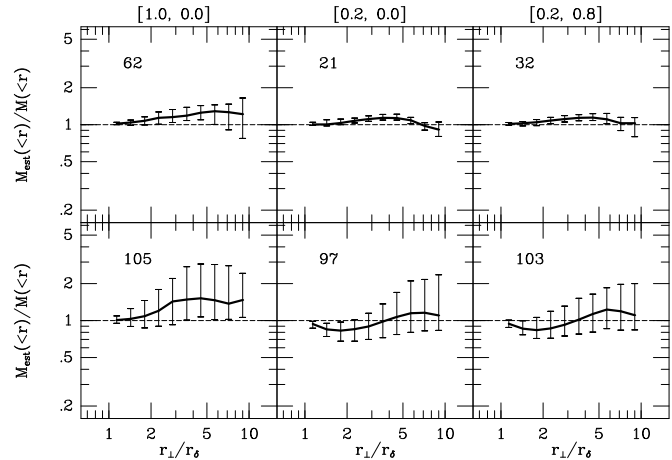}
\caption{Median mass profiles, measured with the caustic technique, 
of dark matter halos in samples extracted from 
CDM models. The cosmological parameters
$[\Omega_0,\Omega_\Lambda]$ are shown above the upper panels.
The upper row shows the most massive halos: $M(<r_\delta)\ge 10^{14} M_{\odot}$ for the 
high-density model, and $M(<r_\delta)\ge 2\cdot 10^{13}M_{\odot}$ for the 
low-density models.
The lower row shows the least massive halos: $10^{13}M_{\odot}\le
M(<r_\delta)< 10^{14} M_{\odot}$ for the high-density model, and
$10^{12}M_{\odot}\le M(<r_\delta)< 2\cdot 10^{13} M_\odot$ for the low-density
models. The numbers of halos in each sample are indicated 
in each panel. The error bars indicate upper and lower
quartiles at each projected distance $r_\perp$. From \cite{diaf97}.}
\label{fig:nbody97}
\end{figure}

\begin{figure}
\centering
\includegraphics[angle=90,width=0.8\textwidth]{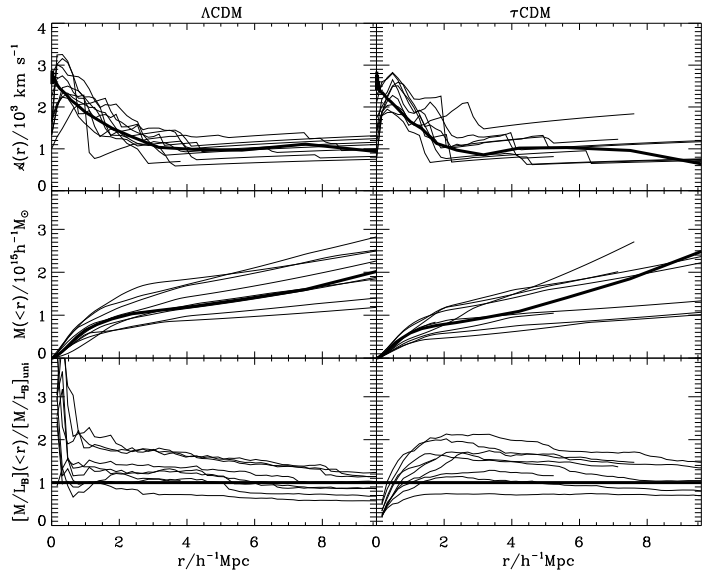}
\caption{Radial profiles of the caustic amplitude (upper row), cumulative
mass (middle row), and mass-to-light ratio (lower row) of a simulated cluster
observed along ten different lines of sight. The thin lines are the
profiles estimated from the individual redshift diagrams. The thick
lines are the real profiles. In the lower panels, the solid line
is the mean mass-to-light ratio of the simulated universe. Left
and right columns are for a cluster in a $\Lambda$CDM and a $\tau$CDM model,
respectively. From \cite{diaf99}.}
\label{fig:nbody99}
\end{figure}

\subsection{Comparison with simulations}

The caustic technique was tested on $N$-body simulations of cluster formation
in CDM cosmologies.
Dark matter only simulations  
showed that the caustic amplitude and the escape velocity profiles
agree amazingly well out to ten virial radii, independently
of the cosmological parameters and, more importantly, of the dynamical state
of the cluster (Figure \ref{fig:vesc}). These simulations also showed that the
technique works on both massive and less
massive clusters (Figure \ref{fig:nbody97}).
In the latter case, the scatter is larger because of projection effects 
and sparse sampling. In the most massive systems, the mass is recovered
 within $20\%$ out to ten virial radii in most cases.

To test the implementation of the caustic method in realistic cases, 
we can use $N$-body simulations where
the galaxies are formed and evolved with a semi-analytic technique \cite{kauffmann99}.
Figure \ref{fig:nbody99} shows the mass profile
of a single cluster observed along ten different lines of sight
in such simulations \cite{diaf99}. 
When comparing this figure with Figures \ref{fig:vesc} and \ref{fig:nbody97},
where all the dark matter particles were observed, we see that the 
caustic technique performs in this latter case better than when 
only the galaxies are available. This difference clearly originates
from the sparser sampling of the velocity field provided by the galaxies.
Figure \ref{fig:nbody99} also shows that projection effects cause the
most relevant systematic error. However, the uncertainty on the mass profile 
remains smaller than 50\% out to $8 h^{-1}$~Mpc from the cluster center.

\begin{figure}
\centering
\includegraphics[angle=90,width=0.75\textwidth]{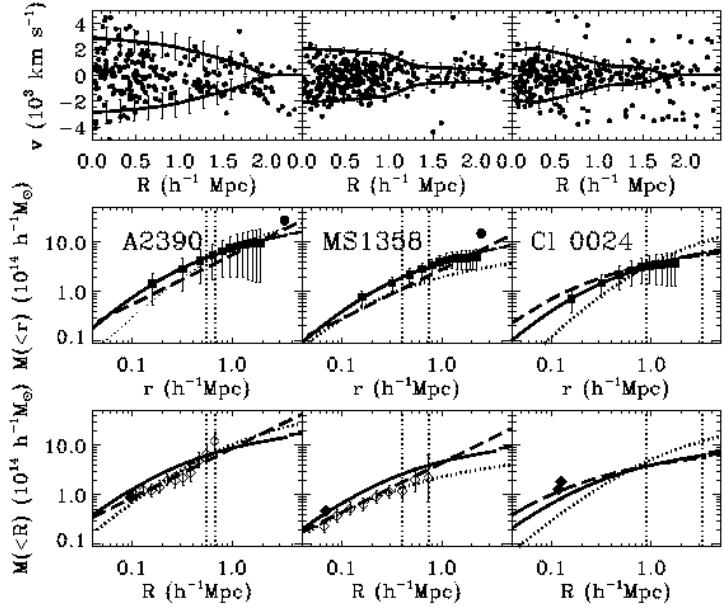}
\caption{Comparison between caustic, lensing, and X-ray mass estimates. The left, middle and right columns are for
A2390, MS1358 and Cl~0024, respectively. {\it Top panels}: Redshift diagrams with the
galaxies (dots) and caustic locations (solid lines). Line-of-sight velocities $v$ are
in the cluster rest-frame. {\it Middle panels}: Three-dimensional cumulative mass profiles. The solid
squares show the caustic mass estimates;
the solid lines are the best-fitting NFW profiles to the data points
within $1 h^{-1}$ Mpc; the dotted lines are the best-fitting NFW profiles to the X-ray measures (from left to right:
\cite{allen01, araba02, ota04}); the dashed
lines are the best-fitting isothermal (A2390,
\cite{squires96}; MS1358, \cite{hoekstra98})  or NFW models  (Cl~0024, \cite{kneib03})
to the gravitational
lensing measures. The left and right vertical dotted lines show the radius of the X-ray
and gravitational lensing fields of view, respectively.
The two filled circles show the virial estimates 
of A2390 and MS1358 \cite{carlberg96}. {\it Bottom panels}:
Projected cumulative mass profiles; lines are as in the middle panels.
The open diamonds show the weak lensing measures: A2390, \cite{squires96};
MS1358, lower limit to the mass profile \cite{hoekstra98}.
Filled diamonds show the strong lensing measures: A2390, \cite{pierre96};
MS1358: \cite{allen98, franx97};
Cl~0024: upper symbol, \cite{tyson98}, lower symbol, \cite{broadh00}.
Error bars in all panels are 1-$\sigma$;
error bars on points where they seem to be missing are smaller than the symbol size.
From \cite{diaf05}.}
\centering
\label{fig:caus-vs-lens}
\end{figure}

\subsection{Caustic vs. lensing}

In equation (\ref{eq:recipe-massprof}) the choice of the constant filling factor ${\cal F}_\beta$ 
is based on $N$-body simulations alone. Therefore, it is not
guaranteed that the caustic technique can recover the mass profile
of real clusters if the simulations are not a realistic representation
of the large-scale mass distribution in the Universe.

Other than the caustic technique, the only method 
for estimating the mass in the outer regions of galaxy clusters is based on weak
lensing. The comparison between these two methods was performed 
on the clusters A2390, MS1358 and Cl~0024 which are at the appropriate 
redshift to have a reasonably intense lens signal and a sufficiently high
number of galaxy redshifts \cite{diaf05}. 
Figure \ref{fig:caus-vs-lens} shows the redshift diagrams and the
mass profiles of these systems. Caustic and lensing masses agree
amazingly well. The most impressive result is for Cl~0024.
This cluster is likely
to have experienced a recent merging event \cite{czoske02}, and 
it probably is out of equilibrium: in this system the caustic mass
and the lensing mass agree with each other, but disagree with the X-ray mass, which
is the only estimate relying on dynamical equilibrium. This result therefore proves
the reliability of the caustic technique and its independence of
the dynamical state of the system in real clusters.

\section{Application to real systems}

\begin{figure}
\centering
\includegraphics[angle=180,width=0.7\textwidth]{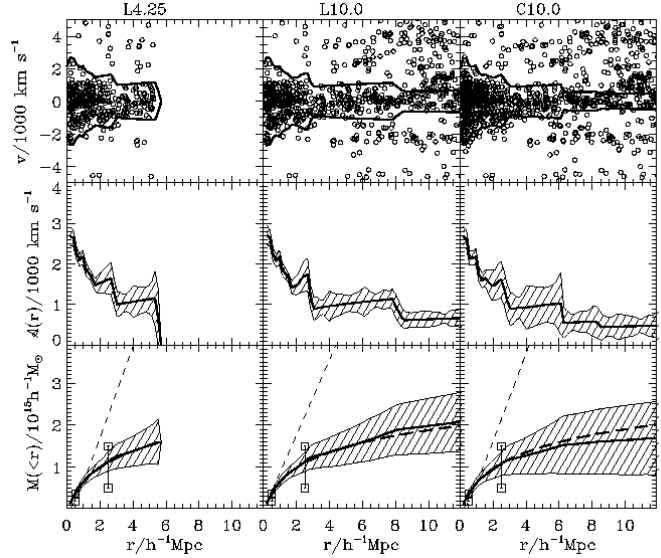}
\caption{{\bf Top panels}: Galaxy distribution in the redshift diagram of Coma for
three galaxy samples of increasing size. 
There are 332, 480, and 691 galaxies within the caustics in the samples
L4.25, L10.0, and C10.0, respectively. Note that these samples are not
substantially larger than the samples in Figure \ref{fig:vanhaarl93}
used to estimate $\Omega_0$ with the spherical infall model.  
The bold lines indicate the location of the caustics. Half the
distance between the caustics defines the amplitude
${\cal A}(r)$ shown in the middle panels.
{\bf Bottom panels}: The bold lines are the caustic mass profiles. 
The two error bars show the range of the
X-ray mass estimates listed in \cite{hughes89}.
Short-dashed and long-dashed lines are the cumulative
mass profile for a softened isothermal sphere and an NFW density profile with
parameters obtained by fitting the mass profile in the range $[0,1]h^{-1}$ Mpc.
Shaded areas in the middle and bottom panels indicate the 2-$\sigma$ uncertainty. From \cite{diaf05}.}
\label{fig:coma}
\end{figure}

\subsection{Mass profiles}

Geller {\it et al.} \cite{geller99} were the first to apply the caustic method to a 
real cluster: they
measured the mass profile of Coma
out to $10 h^{-1}$ Mpc from the cluster center and were able
to demonstrate that the NFW profile fits the cluster
density profile out to these very large radii, thus ruling out the isothermal sphere as
a viable model of the cluster mass distribution (Figure \ref{fig:coma}). 
A few years later, the failure of the isothermal model
was confirmed by the first similar analyses based on gravitational lensing
applied to A1689 \cite{clowe01, lemze08} and Cl~0024 \cite{kneib03}.
The goodness of the NFW fit out to $5-10 h^{-1}$ Mpc was confirmed by applying the caustic technique
to a sample of nine clusters densely sampled in their outer
regions, the Cluster And Infall Region Nearby Survey (CAIRNS, \cite{rines03}), and, more
recently, to a complete sample of 72 X-ray selected clusters with galaxy redshifts
extracted from the Fourth Data Release of the Sloan Digital Sky Survey (Cluster Infall
Regions in the Sloan Digital Sky Survey: CIRS, \cite{rines06}).

\begin{figure}
\centering
\includegraphics[angle=0,width=0.5\textwidth]{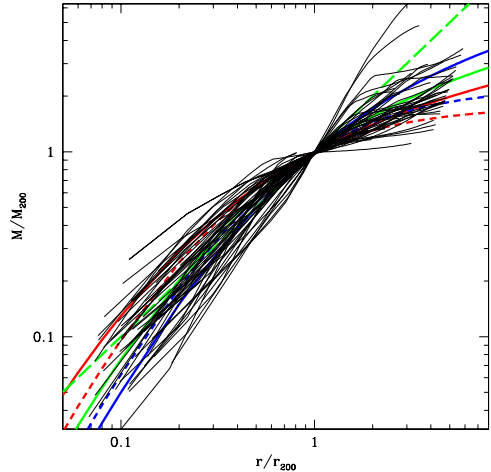}
\caption{Scaled caustic mass profiles for the CIRS
clusters. The thin solid lines show the
caustic mass profiles normalized by $r_{200}$ and $M_{200}$, the total
mass within $r_{200}$.  The
long-dashed line shows a singular isothermal sphere, the solid
lines show NFW profiles (with concentrations $c=3, 5, 10$ from top to
bottom at large radii).  The short-dashed lines are Hernquist profiles
with scale radii different by a factor of two. From \cite{rines06}.}
\label{fig:cirs}
\end{figure}

CIRS is currently the largest sample of clusters whose mass profiles have been measured 
out to $\sim 3 r_{200}$ (Figure \ref{fig:cirs});
Rines and Diaferio \cite{rines06} were thus able to obtain a statistically 
significant estimate of the
ratio between the mass within the turn-around radius 
$M_{\rm t}$ and the virial mass $M_{200}$: 
they found an average value of $M_{\rm t}/M_{200} = 2.2\pm 0.2$,
which is $\sim 50\%$ smaller than the value expected in current models of 
cluster formation \cite{tinker05}. The caustic technique is not limited
to clusters, but, when enough redshifts
are available, it can also be applied to groups
of galaxies: on a sample of 16 groups both the NFW mass profiles and
the ratio $M_{\rm t}/M_{200} = 2.3\pm 0.4$ are confirmed \cite{rines08b}.

 Rines {\it et al.} \cite{rines07, rines08} also
used the CIRS sample to estimate the virial mass function of nearby clusters and determined
cosmological parameters consistent with the WMAP values \cite{dunkley08}; they also showed that
velocity bias is absent in real clusters.

A good fit with the NFW profile out to $\sim 2 r_{200}$ was also found by
Biviano and Girardi \cite{biviano03} 
who applied the caustic technique to an ensemble cluster obtained
by stacking 43 clusters from the Two Degree Galaxy Redshift Survey (2dGFRS, \cite{colless01}):
here, unlike the previous analyses, the
caustic method was not applied to individual clusters, because the
number of galaxies per cluster was relatively small.

The caustic method does not rely on the dynamical state of the cluster and
its external regions: there are therefore
estimates of the mass of unrelaxed systems, for example, among others,
the Shapley supercluster \cite{reisenegger00}, the poor Fornax
cluster, which contains two distinct dynamical components \cite{drinkwater01}, the A2199 complex
\cite{rines02}.

\subsection{Mass-to-light profiles}

By combining accurate photometry with the caustic mass
of A576, Rines {\it et al.} \cite{rines00} were able to measure, for the first time, 
the profile of the mass-to-light ratio $M/L$ 
well beyond the cluster virial radius: they found an $R$-band $M/L$ profile steadily decreasing
from $\sim 0.5$ to $4 h^{-1}$ Mpc, indicating that, in this cluster, dark matter is more concentrated than
galaxies. Slightly decreasing $M/L$ profiles were also measured in the outer region of
five (including A576) out of the nine CAIRNS clusters in the $K$-band \cite{rines04}.
The remaining CAIRNS clusters have an $M/L$ profile which remains roughly flat at radii larger than $\sim 1h^{-1}$ Mpc.
Coma shows a remarkably flat $K$-band $M/L$ profile out to $10h^{-1}$ Mpc \cite{rines01a}.
A flat $M/L$ profile beyond $\sim 0.5 h^{-1}$ Mpc was also found in A1644 in the $H$-band \cite{tustin01}.

These results are due to two reasons: (1) a larger predominance of less luminous late-type galaxies in the
cluster outer regions; and (2) the fact that the $K$-band $M/L$ ratio 
of real galaxy systems increases 
with the system mass \cite{ramella04}. In fact, clusters form by accretion of smaller
systems, as indicated for example by the optical and X-ray observations of the A2199 complex \cite{rines01b},
and as expected in current hierarchical models of structure formation \cite{springel06}; therefore,
the cluster surrounding regions, which mostly contain galaxy groups,
should naturally have smaller $M/L$. The positive $M/L-$mass correlation was also obtained
in semi-analytical models of galaxy formation \cite{kauffmann99} and is well
described by the statistical technique based on the conditional luminosity
function \cite{vandenbosch04}.

The infall regions are the transition between the dense cluster regions
and the field \cite{balogh04, IAU195}, and the internal properties of galaxies do not vary abruptly at the
virial radius \cite{rines05}. Therefore galaxy surveys in 
the outskirts of clusters, as those mentioned above, can clearly
constrain models of cluster and galaxy formation. 

\section{Conclusion and perspectives}

The caustic method and gravitational lensing are the
only two techniques currently available for measuring the mass profile
of clusters beyond their virial radius. 
The caustic method requires a sufficiently dense redshift survey with a large field of view 
and is only limited by the time needed to 
measure a large enough number of galaxy spectra; this observing time 
 increases quickly with cluster redshift.
On the other hand, lensing requires wide-field photometric surveys that need 
high angular resolution and extremely good observing conditions; moreover, the lensing 
signal is strong enough only when the cluster is within a limited
redshift range $z\approx 0.1-1$.  

When the caustic technique was proposed, multi-object spectroscopy
was not routinely applied to measure galaxy redshifts, and the request of
100 or more redshifts in the outskirts of clusters appeared demanding.
Nowadays this task can be accomplished more easily and the popularity of 
the caustic technique has begun to increase.

The caustic technique has been tested on $N$-body simulations and the 
mass profiles are accurate to better than $\sim 50\%$ out to $\sim 3-4 r_{200}$.
On the three systems where both the caustic method and lensing
could be applied, the two methods yield consistent mass profiles.
This consistency also holds in Cl~0024 whose X-ray mass profile disagrees
with the caustic and lensing profiles; this disagreement is most likely 
due to the fact that
this cluster is out of equilibrium and thus the X-ray mass is unreliable.

The uncertainties on the caustic mass profile are almost totally due
to projection effects. In fact, the method assumes that the cluster is 
spherically symmetric, and this is rarely the case; therefore the redshift diagram
from which the caustic mass is extracted can vary substantially when
the cluster is observed along different lines of sight.
The size of this systematic error is comparable to the systematic uncertainty
we have with lensing methods which measure
all the mass projected along the line of sight. 

What the caustic technique actually measures is the line-of-sight
component of the escape velocity from the cluster (equation \ref{eq:rig-pot}). If we
can measure the velocity anisotropy parameter $\beta$, the caustic
technique thus yields a direct measure of the profile of the cluster gravitational potential.
 
This brief review shows that the caustic technique is a powerful tool
for the analysis of clusters and their external regions, but its full potentiality 
still needs to be exploited. For example, the $\sigma$ plateau,
that appears when walking along the binary tree (Figure \ref{fig:thresholds}), 
provides a clean way to identify the cluster members. 
This issue still needs a throughout investigation \cite{serra},
but very preliminary results, based on a large sample of synthetic clusters,
show that $\sim 90\%$ of the galaxies within the caustics
are cluster members and that the interloper contamination is comparable
or lower than other methods \cite{wojtak07}.
An additional byproduct of the caustic machinery is the identification 
of cluster substructures from the distribution of the galaxies
in the binary tree \cite{serna96}. This topic has also been currently investigated \cite{serra}.

\acknowledgments
I thank Alfonso Cavaliere and Yoel Rephaeli for the invitation
to this fruitful and well organized school. It is a pleasure to acknowledge the hospitality
of SIF during my stay in Varenna. I wish to thank 
Margaret Geller and Ken Rines who largely contributed to the
development and dissemination of the caustic method. I also thank them for a
careful reading of the manuscript and suggestions. 
Support from the PRIN2006 grant ``Costituenti fondamentali dell'Universo'' of
the Italian Ministry of University and Scientific Research
and from the INFN grant PD51 is gratefully acknowledged.

\end{document}